# Lattice dynamics and spin-phonon interaction in strained NiO films

Alireza Kashir[1]*, Veronica Goian[2], Daseob Yoon[3], Byeong-Gwan Cho[4], Yoon Hee Jeong[1, 5], Gil-Ho Lee[1], Stanislav Kamba[2]**

[1]Department of Physics, Pohang University of Science and Technology (POSTECH), Pohang, 37673, Republic of Korea
[2]Institute of Physics, Czech Academy of Sciences, Na Slovance 2, 182 21 Prague 8, Czech Republic
[3]Department of Materials Science and Engineering, Pohang University of Science and Technology (POSTECH), Pohang, 37673, Republic of Korea
[4]Pohang Accelerator Laboratory, Pohang, 37673, Republic of Korea
[5]Department of Physics, Korea Advanced Institute of Science and Technology, Daejon, 34141, Republic of Korea

## Abstract

NiO thin films with various strains were grown on $SrTiO_3$ (STO) and MgO substrates using a pulsed laser deposition technique. The films were characterized using an X-ray diffractometer, atomic force microscopy, and infrared reflectance spectroscopy. The films grown on STO (001) substrate show a compressive in-plane strain which increases as the film thickness is reduced resulting in an increase of the NiO phonon frequency. On the other hand, a tensile strain was detected in the NiO film grown on MgO (001) substrate which induces a softening of the phonon frequency. Overall, the variation of in-plane strain from -0.36% (compressive) to 0.48% (tensile) yields the decrease of the phonon frequency from 409.6 $cm^{-1}$ to 377.5 $cm^{-1}$ which occurs due to the ~ 1% change of interatomic distances. The magnetic exchange-driven phonon splitting $\Delta\omega$ in three different samples, with relaxed (i.e. zero) strain, 0.36% compressive strain and 0.48% tensile strain, was measured as a function of temperature. The $\Delta\omega$ increases on cooling in NiO relaxed film as in the previously published work on a bulk crystal. The splitting increases on cooling also in 0.48% tensile strained film, but $\Delta\omega$ is systematically 3-4 $cm^{-1}$ smaller than in relaxed film. Since the phonon splitting is proportional to the non-dominant magnetic exchange interaction $J_1$, the reduction of phonon splitting in tensile-strained film was explained by a diminishing of $J_1$ with lattice expansion. Increase of $\Delta\omega$ on cooling can be also explained by rising of $J_1$ with reduced temperature.

E-mails: *kashir@postech.ac.kr

**kamba@fzu.cz

## Introduction

Cubic magnetic binary monoxides with rocksalt crystal structure (Fig. 1, space group $Fm\bar{3}m$) belong to a group of insulating systems with antiferromagnetic (AFM) spin structure through superexchange magnetic interaction [1]. The simplicity of their crystal and electronic structure draws huge attention from theorists as a model system to investigate the emergent physical phenomena which appear under complex condition e.g. spin-phonon coupling [2-4], magnetic ordering [5] and a high-pressure insulator-metal transition [6-8]. Recently, the effect of strain on the electronic, structural and phonon properties of rocksalt magnetic binary compounds became one of the challenging topics in solid-state physics. Induction of ferroelectricity [9, 10] and even multiferroicity [11-13] has been observed. By applying a proper strain, the spin-phonon interaction [14] metal-insulator transition [6-7], magnetic moments [8] and dielectric permittivity [9, 15-16], can be tuned. This was theoretically predicted and experimentally approved by several research groups. During the last decade, the study of the influence of strain on the lattice dynamics of *rocksalt binary oxides* has become the researchers' primary focus, as there exists a direct connection between structural features and lattice dynamical behavior.

Bousquet *et al.* [9] and Liu *et al*. [12], using first-principles density functional calculations, and Kim [10] by applying a combination of *ab initio* calculation and soft mode group theory analysis studied the effect of epitaxial strain on the lattice dynamics of various rocksalt binary compounds. It was predicted that the epitaxial strain can lower cubic symmetry ($Fm\bar{3}m$) to tetragonal (*I4/mmm*) and finally to an orthorhombic ferroelectric phase. The latter structural phase transition should be driven by an optical soft phonon. The phonon softening should cause an increase of dielectric permittivity $\varepsilon'$ in the films due to the Lyddane-Sachs-Teller relation [17]. Besides many diamagnetic binary compounds, their approaches were applied also to the ferromagnetic EuO [9] and GdN [12], and surprisingly it was predicted that under proper epitaxial strain both systems should exhibit a ferroelectric transition in their preserved ferromagnetic ground state, i.e. the materials should become multiferroic. Recently, we showed that the low-temperature dielectric permittivity of pulsed laser deposited EuO [15] and NiO [16] thin films increases with the strain and EuO even exhibits a proper (and displacive) ferroelectric phase transition, if the tensile strain exceeds 6% [13]. Wan *et al*. [11] theoretically predicted a spin-induced (i.e. improper) ferroelectric phase transition in AFM MnO, if the strain exceeds 4%. Their calculation shows that more than 4% compressive strain leads to an off-center shift of oxygen anion in MnO due to the magnetic exchange interaction resulting in a non-centrosymmetric tetragonal structure.

Recently, we studied the effect of strain on the spin-phonon interaction in MnO thin films and observed a substantial increase of the phonon splitting $\Delta\omega$ in the AFM phase under a compressive biaxial strain [14]. According to Luo *et al*. [18], the phonon splitting occurs due to nearest neighbor (nn) exchange interaction between transition-metal cations (described by coupling constant $J_1$), while the AFM order is controlled by next nearest neighbor (nnn) coupling constant $J_2$ from the Heisenberg Hamiltonian:

$$H = \sum_{nn} J_1 S_i S_j + \sum_{nnn} J_2 S_i S_j \qquad (1)$$

Fischer *et al.* [19] calculated $J_1$ in several transition metal monoxides as a function of lattice constants and predicted a substantial change of $J_1$ with the pressure or strain (Fig. 2(a)). The prediction was later independently confirmed in [20, 21]. Kant *et al.* [22] even discovered a linear relation between $\Delta\omega$ and coupling constant $J_1$ in various transition-metal monoxides and Cr-spinels. Based on this linear relation and the pressure dependence of $J_1$, one can explain the observed increase of $\Delta\omega$ in compressively strained MnO films [14].

Almost all transition-metal monoxides have $J_1 > 0$, but NiO has $J_1 < 0$ [18] and subsequently its $\Delta\omega = \omega_{||} - \omega_{\perp} < 0$ [22] ($\omega_{||}$ and $\omega_{\perp}$ mark phonon frequencies along and perpendicular to the AFM axis – see Fig. 2(b)). Moreover, it was predicted that $J_1$ in NiO should exhibit only small change with the strain (Fig. 2(a) and [19]). The effect of strain on the lattice dynamics of NiO and its comparison with MnO are subject of this article, because they can confirm or exclude Fischer's theoretical predictions [19] and provide more information about origin of the phonon splitting in AFM rocksalt transition metal monoxides.

Nickel oxide is the only stable oxide in Ni-O phase diagram [23] crystallizing in high-temperature paramagnetic phase in cubic rocksalt structure with the lattice parameter of 4.177 Å [1] ($Fm\bar{3}m$) (Fig. 1). $Ni^{2+}$ cations are located in octahedral environments constructed by $O^{2-}$ anions resulting in the split of fivefold degenerate $d$-electron-levels into a lower $t_{2g}$ triplet and an upper $e_g$ doublet with an insignificant spin-orbit coupling [4].

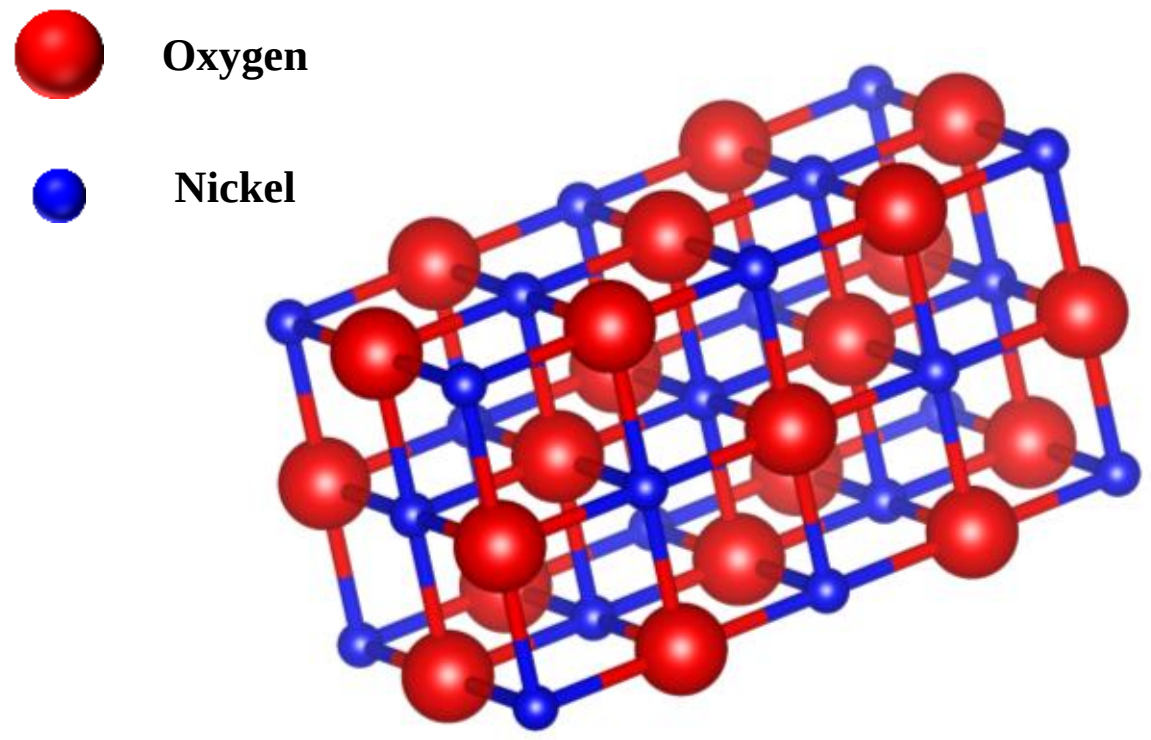

**Fig. 1.** Crystal structure of cubic NiO above $T_N$.

NiO belongs to the wide band-gap (~ 4 eV) type-II antiferromagnets (AFII) with the Néel temperature $T_N$ of 523 K [1]. In AFII state, the magnetic cations located in {111} planes align ferromagnetically in $[11\bar{2}]$ direction. These ferromagnetic sheets are antiferromagnetically arranged in <111> direction (Fig. 2) [22]. Accompanying the AFII magnetic transition, exchange-striction causes a small structural distortion at which the high-temperature cubic rocksalt structure becomes rhombohedrally distorted along the <111> direction which slightly compresses the unit cell and the angle between the lattice vectors increases to 90.07 ° ($R\bar{3}m$, No. 166) [24, 25].

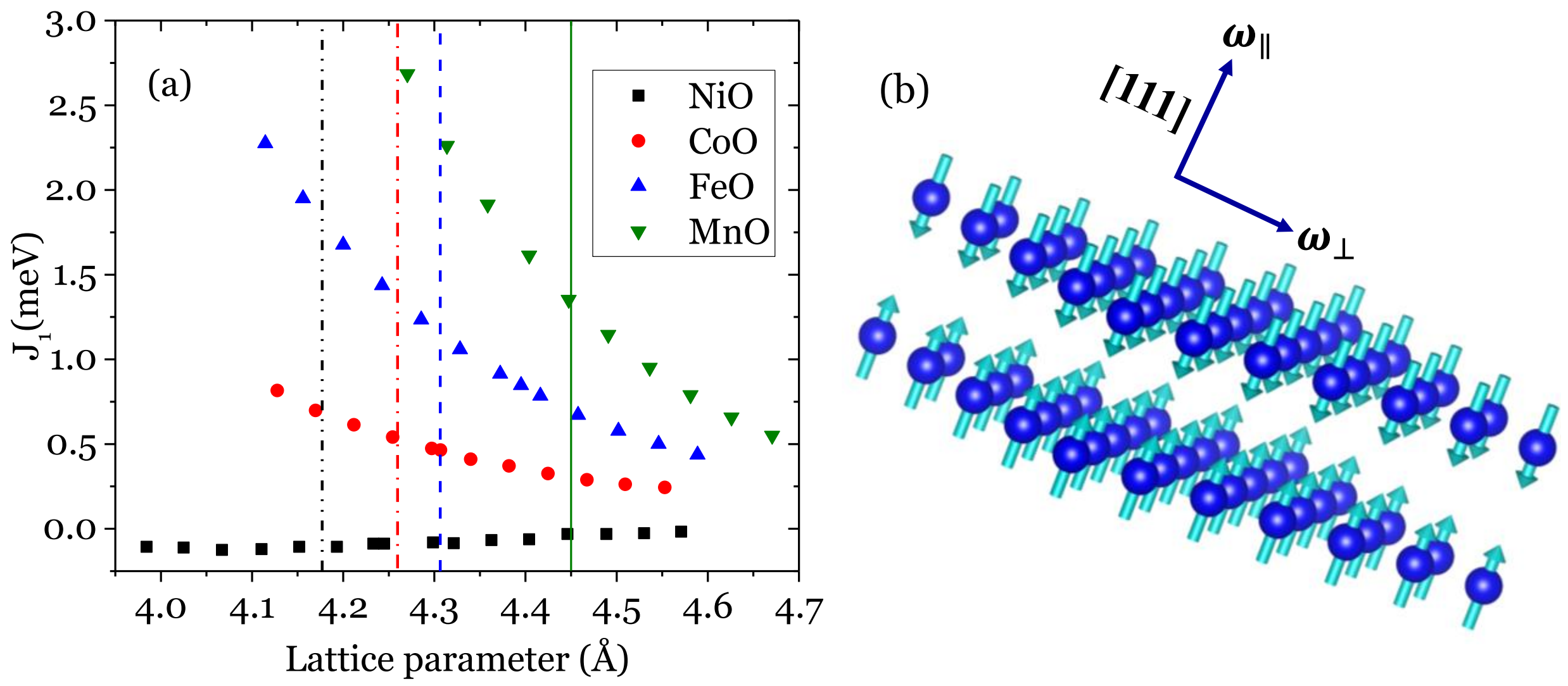

**Fig. 2.** (a) The $J_1$ dependency on the lattice constant for various transition metal monoxides calculated by Fischer *et al.* [19]. The sign of $J_1$ was reversed in agreement with Ref. [22]. The vertical lines mark experimental lattice constants in crystals, adapted from [19]. (b) Schema of antiferromagnetic order along [111] direction in NiO and exchange-driven phonon splitting into two eigenfrequencies parallel $\omega_{\parallel}$ and perpendicular $\omega_{\perp}$ to the AFM order. Oxygen atoms are not displayed for simplicity.

The lattice dynamics of nickel oxide have been studied by several theorists and experimentalists [4, 18, 22, 24-28]. In the paramagnetic phase, NiO has a single triple degenerate transverse optical (TO) phonon in the $\Gamma$ point with $F_{1u}$ symmetry and frequency near 384 $cm^{-1}$ [4]. Below $T_N$ = 523 K the phonon splits into two different modes with eigenfrequencies $\omega_{\parallel}$ and $\omega_{\perp}$ polarized parallel and perpendicular to the AFM direction, respectively (Fig. 2) [22]. Uchiyama *et al.* [26] studied the lattice dynamics of NiO at RT using inelastic X-ray scattering. They found $\omega_{||} - \omega_{\perp} \sim$ -1.0 meV ($\Delta\omega$ ~ -8 $cm^{-1}$) in the vicinity of the Brillouin zone center. Chung *et al.* [24] investigated the phonon splitting in MnO and NiO using inelastic neutron scattering. Below $T_N$, they observed a phonon splitting of approximately 10% in both MnO and NiO (i.e. twice larger than those in [25]), however, the NiO data are much less conclusive than that for MnO, because inelastic neutron or x-ray scattering have low accuracy. Nevertheless, more precise IR reflectivity studies [4] of bulk NiO crystals confirmed $\Delta\omega \cong 10$ $cm^{-1}$ (at 5 K), which is only 2.5 % of $\omega_{\perp}$. In principle, lattice distortion accompanying the magnetic phase transition can explain the splitting of the phonons due to the reduced lattice symmetry. However, on the basis of local spin density approximations (LSDA) it was clearly stated that the phonon anisotropy can be explained solely by the Born-effective-charge Z* redistribution from spherical (for cubic structure) to cylindrical (for rhombohedral structure) with the AFM axis <111> being the symmetry axis [27]. Although considering the effect of small structural distortion, the splitting is predictable, but the value of splitting would be one order of magnitude smaller than that of the observed one [25]. Moreover, in case of NiO, below $T_N$, due to the lattice compression along <111> direction [18], the phonon mode parallel to the <111> direction $\omega_{\parallel}$ is supposed to have a higher frequency compared to the perpendicular one $\omega_{\perp}$. However, $\Delta\omega = \omega_{\parallel} - \omega_{\perp} < 0$ [22, 25, 28]. Theoretical studies attempted to explain the nature of spin-phonon coupling in detail [18, 25, 28]. Recently, Luo *et al.* [18] proposed for the rocksalt magnetic binary oxides that the size of the exchange-induced phonon splitting $\Delta\omega$ is *solely* determined by the sign and magnitude of the non-dominant exchange constant $J_1$. There is a linear relation between $\Delta\omega$ and $J_1$ and it can be described using the formula $\hbar\Delta\omega = \beta J_1 S^2$ [22]. The dominant 180° superexchange coupling $J_2$ has no contribution to the actual splitting $\Delta\omega$. NiO has small and negative $J_1$ (Fig. 2(a)) which proposes a negative sign for $\Delta\omega$ in agreement with the experimental result. The theoretical prediction was experimentally confirmed by Park and Choi [28] and Kant *et al.* [22] found that the negative sign of splitting in the case of NiO is directly related to the negative sign of $J_1$. Therefore, besides the direct effect of strain, the behavior of nearest-neighbor exchange interaction $J_1$ under strain controls the lattice dynamics of NiO below $T_N$. Here, using infrared spectroscopy, we investigate the effect of strain and temperature on the lattice dynamics of NiO thin films grown on different substrates and compare the results with phonons in strained MnO films.

## Experiments

### Substrate preparation

$SrTiO_3$ (001) (STO) and MgO (001) substrates (from Crystech) were used to grow NiO films with different thicknesses and strain. Before deposition, all substrates were ultrasonically cleaned by acetone and methanol for 10 min to remove any possible contaminants from their top surface. STO substrate was passed through a chemical etching process; first for 5 min agitation in de-ionized water and then a 30-second soaking in buffered hydrogen fluoride ($NH_4F$ : HF=7 : 1) (pH~ 4.5) to achieve a single terminated $TiO_2$ surface. Then the annealing processes in the air were done on substrates; 2.5 hours at 1000 °C for STO and 3 hours at 1150 °C for MgO. These annealing processes provide a single terminated step and terrace surface structure for both kinds of substrate [29, 30].

### Film growth

The pulsed laser deposition technique (PLD) was used to prepare all thin films in this work. A KrF laser ($\lambda_{laser}$ = 248 nm) operating at 10 Hz generated the laser beam with the energy of 50 mJ per pulse. Through an optical lens, the laser beam was focused on the rotating NiO polycrystalline pellet making a 1×2 mm$^2$ spot size to ablate target materials. The PLD chamber was pumped to 13.3×10$^{-7}$ Pa, then it was filled with 2 Pa high pure $O_2$ gas during the deposition to compensate for any possible oxygen deficiencies in the deposited films. To grow NiO films with different levels of strain, the substrate temperature $T_G$ was changed after each deposition [29]. After growth, the chamber was cooled to the room temperature RT with a rate of 10 °C/min under the growth condition. NiO films with different levels of thicknesses $t_F$ and strain were prepared for IR spectroscopy.

### Structural Characterization

The phases of the deposited films and their crystal quality were evaluated using an X-ray diffractometer with a copper source operating at 40 V and 200 mA, followed by a $\omega$ rocking scan around the detected NiO Bragg peaks. The lattice parameters were determined from lattice spacing of (002) and (204) Bragg planes by using high-resolution X-ray diffraction (HRXRD, Bruker D8 Discovery X-ray diffractometer) with Cu Kα1 radiation ($\lambda$ = 1.5406 Å). We measured (002) and (204) peaks because the measurable peaks among commonly allowed Bragg peaks between perovskite (substrate) and rocksalt (film) are only (002) and (204) peaks. All other peaks are out of range or allowed for only one structure. The surface morphologies of the substrates and the deposited samples were investigated using atomic force microscopy in dynamic non-contact mode on a Park system XE100 scanning probe microscope. An X-ray reflectometer (XRR) was used to measure film thickness $t_F$ and the quality of surface and of the film / substrate interface. To determine $t_F$, we considered the first five oscillations in the XRR pattern and used equation (1) to obtain four different values.

$$t_F \sim \frac{\lambda}{2}\frac{1}{\theta_{m+1}-\theta_m}, \tag{2}$$

where $\lambda$ is the wavelength of X-ray, $\theta_{m+1}$ and $\theta_m$ are the position of *(m+1)-th* and *m-th* interference maxima, respectively. Final $t_F$ was obtained by taking the average of these values and then rounding it to the *nanometer* scale.

**IR Spectroscopy**

The IR reflectivity measurements were performed using a Bruker IFS 113v Fourier transform IR spectrometer equipped with a helium-cooled (1.6 K) silicon bolometer. Spectra were measured at temperatures from 15 to 300 K, which were maintained using an Oxford Optistat CF cryostat with 3-mm-thick polyethylene windows transparent up to 650 cm$^{-1}$. Room temperature (RT) spectra were taken up to 4000 cm$^{-1}$. Polarized and unpolarized reflectance measurements were performed in near-normal incidence geometry.

The reflectance spectra of the thin films and their substrates were evaluated as a two-layer optical system [31]. The reflectivity of the bulk substrate is related to complex relative permittivity $\varepsilon^*(\omega)$ by

$$\boldsymbol{R}(\omega) = \left| \frac{\sqrt{\varepsilon^*} - 1}{\sqrt{\varepsilon^*} + 1} \right|^2 \qquad (3)$$

The $\varepsilon^*(\omega)$ of the substrate is described by a generalized, factorized damped harmonic oscillator model

$$\varepsilon^*(\omega) = \varepsilon_\infty \prod_j \frac{\omega_{LOj}^2 - \omega^2 + i\omega\gamma_{LOj}}{\omega_{TOj}^2 - \omega^2 + i\omega\gamma_{TOj}}, \qquad (4)$$

where $\omega_{\mathrm{TOj}}$ and $\omega_{\mathrm{LOj}}$ are the frequencies of *j*-th transverse optical (TO) and longitudinal optical (LO) phonons, and $\gamma_{\mathrm{TOj}}$ and $\gamma_{\mathrm{LOj}}$ are corresponding damping constants, respectively. $\varepsilon_\infty$ is the high-frequency permittivity (electronic contribution), determined from the room-temperature frequency-independent reflectivity tail above the phonon frequencies. The dielectric function of the thin films was fitted with the sum of *N* independent three-parameter damped harmonic oscillators (representing the in-plane polarized TO phonons of the film), which are expressed as

$$\varepsilon^*(\omega) = \varepsilon_\infty + \sum_{j=1}^{N} \frac{\Delta\varepsilon_j \omega_{TOj}^2}{\omega_{TOj}^2 - \omega^2 + \boldsymbol{i}\omega\gamma_{TOj}} \qquad (5)$$

where $\Delta\varepsilon_j$ is the dielectric strength of the *j*-th mode. We used the equation (5) for fitting of the films because the damping of the LO phonons of the films does not influence the reflectance spectra appreciably and therefore the formula (4) with more free parameters was not necessary to apply.

## Results and Discussion

### Structural Characterization

Atomic force microscopy scans revealed steps of atomic-scale height and flat terraces on the surface of STO and MgO treated substrates (Fig. 3). A line scan (inset of fig. 3(a)) showed steps with an average height of ~ 4 Å, which is the distance between two $TiO_2$ layers along [001] direction in the $SrTiO_3$ cubic perovskite structure and indicated a single terminated surface which was achieved in previous works [29, 30]. A ~ 2.3 Å step height was detected on the surface of the MgO substrate (inset of fig. 3(b)) which is almost the same as the distance between two MgO (001) monolayers.

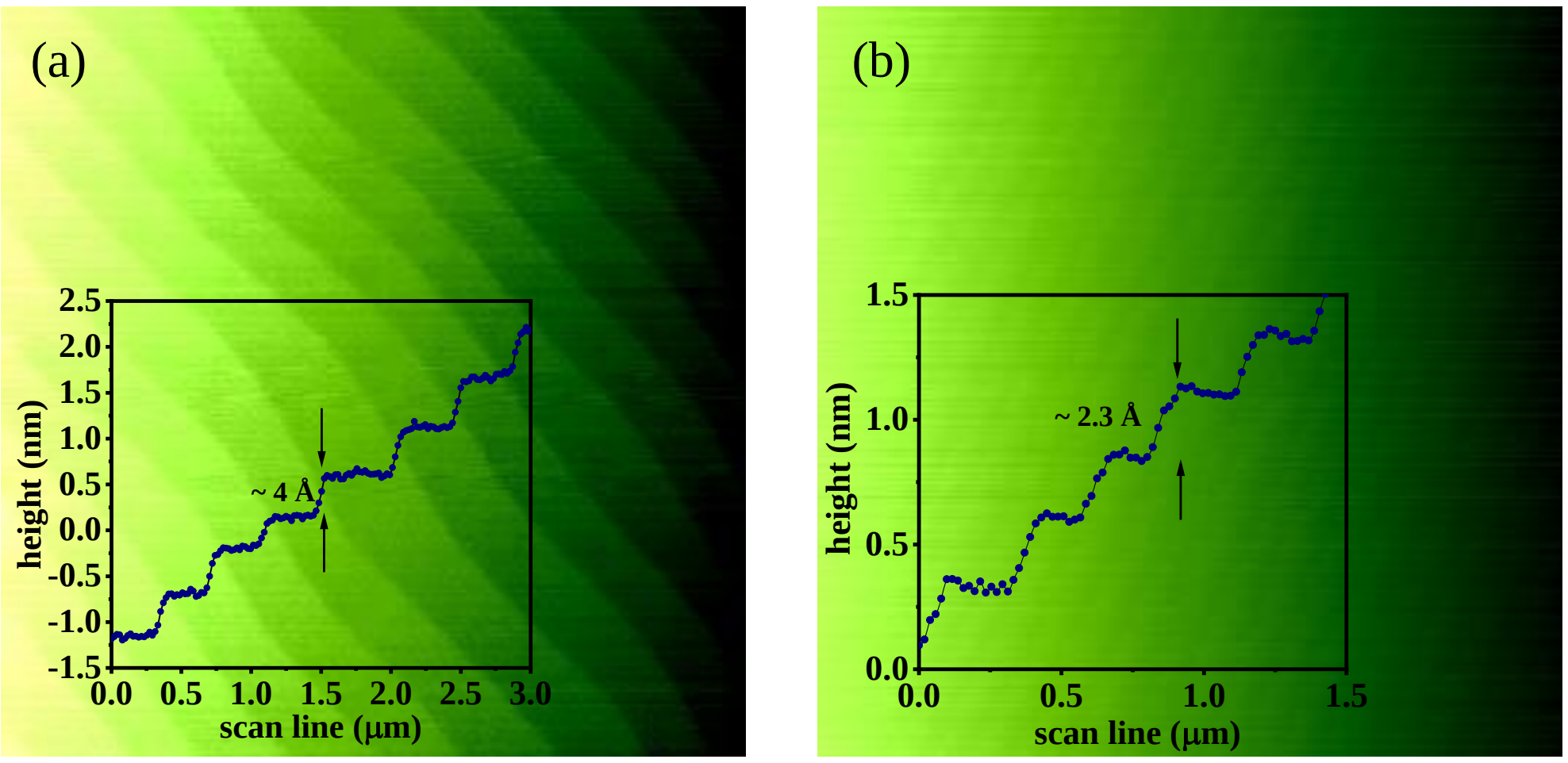

**Fig. 3.** Atomic force microscopy 5×5µm$^2$ topographic images of the substrates' surface after heat treatment. (a) $SrTiO_3$ (001), (b) MgO (001). The insets show line scans on the surface of the substrates.

The XRD patterns of the films deposited on different substrates and at different temperatures are shown in fig. 4. There are seen only NiO and substrate peaks (002) in these scans indicating the absence of any other phases or impurities. The thickest film (~ 71 nm) was fully relaxed and the lattice parameters calculated from the (002) and (204) Bragg peak positions were the same as in the bulk NiO (4.177 Å ). As it was mentioned in our previous work, the strain accommodation in NiO films is a function of $T_G$ and $t_F$ [29]. Decreasing $t_F$ and $T_G$ was accompanied by a left-shift of the NiO (002) peak grown on the STO substrate while a right-shift was observed in the films grown on MgO (002) substrate indicating the different sign of strain accommodated into the NiO structure. Therefore, by tuning the growth condition and selecting proper substrates, we obtained high-quality single-phase NiO films with negative (compressive) and positive (tensile) in-plane strain to study the lattice dynamics of NiO under different strains (Table 1).

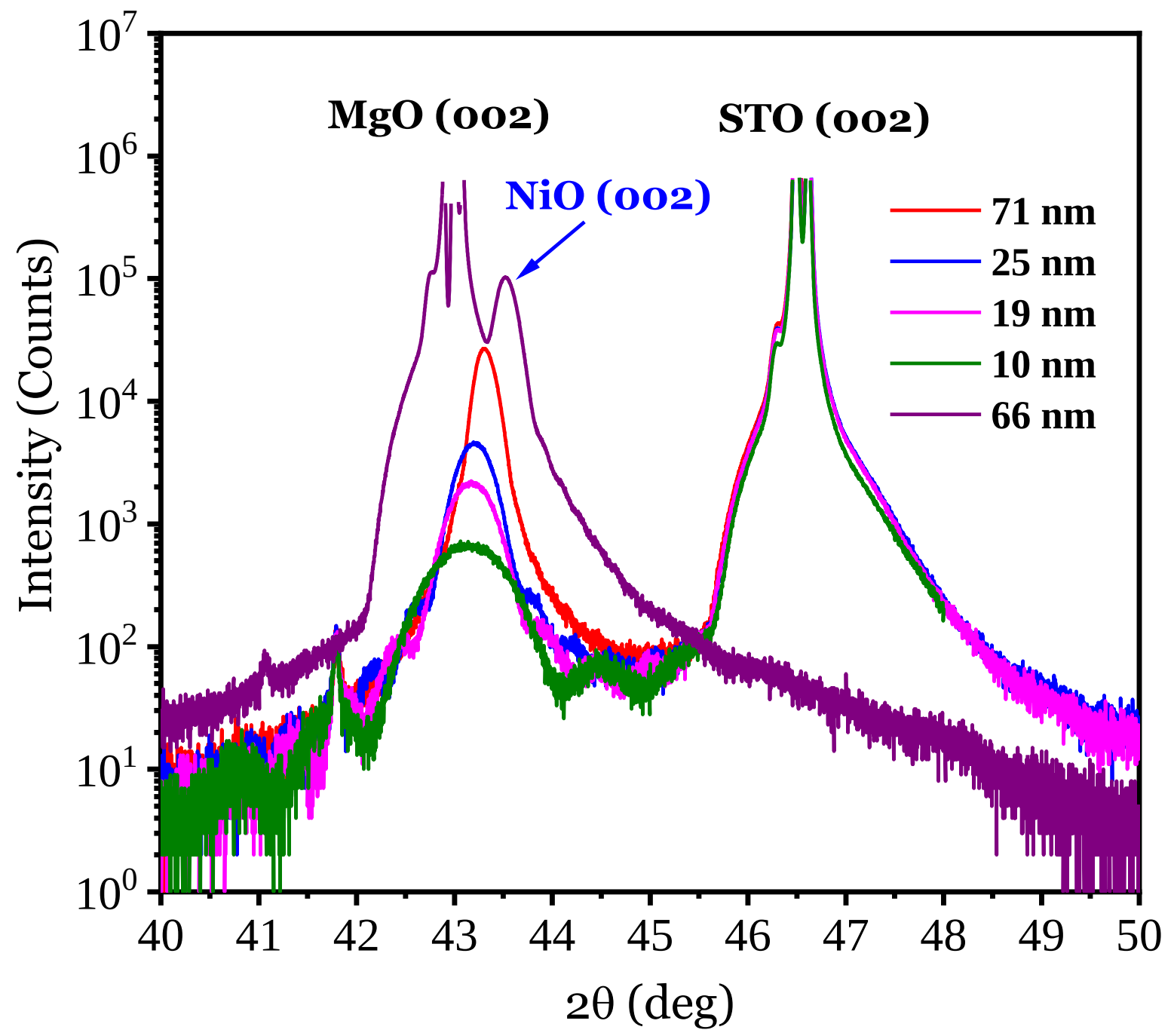

**Fig. 4.** XRD patterns of NiO films grown on STO (001) and MgO (001) substrates, with different thicknesses.

**Table. 1.** Thicknesses and strain level in NiO deposited films.

| Sample's ID | Substrate | Film Thickness | In-plane strain |
|---|---|---|---|
| **A** | MgO (001) | 66 nm | 0.48% |
| **B** | $SrTiO_3$ (001) | 71 nm | 0.0% |
| **C** | $SrTiO_3$ (001) | 25 nm | -0.19% |
| **D** | $SrTiO_3$ (001) | 19 nm | -0.26% |
| **E** | $SrTiO_3$ (001) | 10 nm | -0.36% |

Atomic force microscopy scans of the grown films showed an atomically smooth surface for all films (with different thicknesses) deposited on different substrates (Fig. 5). For the films thinner than 30 nm, the surface consisted of step-and-terrace structure like on the bare substrates while on the films thicker than 50 nm the clarity of these steps faded out and an atomically smooth surface with the average height fluctuation lower than 5 Å was obtained. As the IR reflectivity depends critically on the samples' surface quality, achieving the high-quality surface structure NiO films enabled us to detect the reflectivity spectrum even for films as thin as 10 nm.

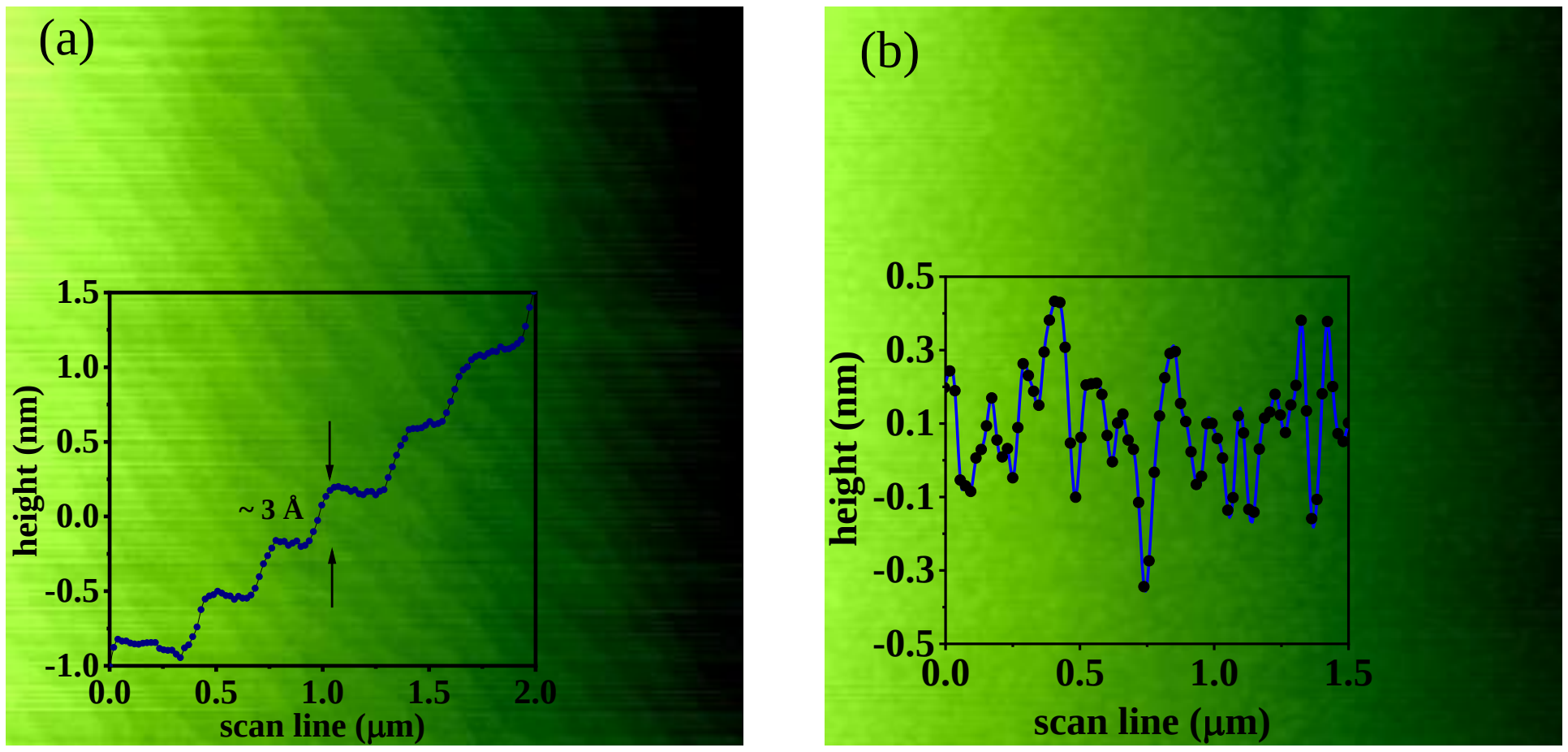

**Fig. 5.** Atomic force microscopy topographic images of NiO thin films deposited on the STO (001) substrate. Films have thicknesses of (a) 25 nm and (b) 71 nm. The insets show line scans on the surface of the samples.

## IR reflectivity spectra

To determine the effect of strain on the lattice dynamics of NiO, the reflectance spectra of the films with various strains were measured. It should be noted that the near-normal reflectivity gives only an *in-plane* phonon response, so we cannot see the phonons polarized perpendicularly to the film plane, i.e. in the [001] direction. If the phonon of the thin film has a frequency in the reststrahlband (i.e. in the frequency range between $\omega_{TO}$ and $\omega_{LO}$) of the substrate, its response is seen as a minimum in the reflectance spectra (Fig. 6(a)). The frequency of the minimum reflectivity is close to the TO phonon frequency $\omega_{TO}$ of the film provided that the dielectric strength of the mode ($f_j$ $=\Delta\varepsilon_j\omega^2_{TOj}$) is low. Nevertheless, we observed from the model fits of made-up thin film with thickness 25 nm that the frequencies of the reflectance minima shift down, if $\Delta\varepsilon_j$ of phonon in the film increases and $\omega_{TOj}$ and $\gamma_{TOj}$ remain constant. This behavior was demonstrated in the inset of fig. 6(a), where the shift of reflectivity minima from 431 to 426 cm$^{-1}$ was seen, if Δε of the model phonon increases from 3 to 12.6 and rest of its parameters does not change (parameters of the phonon are listed in inset of fig. 6(a)).

The Néel temperature $T_N$ of NiO is 523 K, therefore, at RT, we expect, in analogy with reference [4], the splitting of TO phonon mode into two components $\omega_{\parallel}$ and $\omega_{\perp}$, which is driven by the magnetic exchange interaction $J_1$. The NiO lower-energy mode $\omega_{\parallel}$ is polarized parallel to <111> direction and it is hardly observable at RT even in case of bulk NiO [4, 22], although its $T_N$=523

K. The reason is its very weak intensity. In our case, it appears as a low-frequency shoulder of the reflectivity minimum seen near 400 $cm^{-1}$ in figs. 6(a), 7(a) and 7(b), which is determined mainly by the strong double-degenerate $\omega_{\perp}$ mode polarized perpendicular to <111> direction (Fig. 2). The mode is the best observed at 5 K when it has the lowest damping [4, 22]. In our work, it is best seen in the thickest film (Fig. 7(a)). Increasing compressive strain in the films with reduced thickness is accompanied by a reduced mode intensity and the right-shift of minima which indicates an increase (hardening) of the TO phonon frequency (Fig. 6(c)). Qualitatively, Bousquet *et al*. [9] and Kim [10] predicted the same strain dependence of phonon frequencies for in-plane polarized phonons in other rocksalt monoxides. If the phonon frequency of the film is smaller than the TO phonon frequency of the substrate, it is seen as a peak in reflectance spectra (Fig. 6(b)) [32]. The reflectance spectrum of tensile strained NiO film deposited on the MgO substrate shows a small peak at 377.5 $cm^{-1}$, which corresponds to the NiO TO phonon frequency (Fig. 6(b)). Its frequency is approximately 19 $cm^{-1}$ lower than that of the relaxed film (396.9 $cm^{-1}$), which indicates the role of tensile strain in the softening of the phonon. Fig. 6(c) shows the frequency of high-energy mode $\omega_{\perp}$ as a function of the strain. One can see a gradual increase of the $\omega_{\perp}$ frequency from 377.5 to 409.6 $cm^{-1}$ (i.e. 32 $cm^{-1}$ hardening) while the in-plane strain changes from tensile 0.48% to compressive 0.36%. This behavior is a consequence of the increase of electron overlapping between anions and cations located in the (111) planes, resulting in a strong bonding between them. This is in qualitative agreement with the theoretical studies by Bousquet *et al.* [9] and Kim [10] on other binary compounds with the rocksalt crystal structure. According to their calculations, the in-plane polarized phonons with $E_u$ symmetry strongly reduce their frequencies under a tensile strain until they become negative (i.e. the tetragonal crystal structure becomes unstable) at a critical strain. On the other hand, under compressive strain, the frequency of the phonon increases only smoothly with the rising strain [9, 10]. In the range of strain, which we achieved, a *linear* relation was detected between the phonon frequency and atomic distance (Fig. 6(c)) which is in good qualitative agreement with the theoretical predictions [9, 10]. Nevertheless, here should be stressed that the phonon frequency is relatively much higher in NiO than in other binary compounds with rocksalt crystal structure, so the critical strain for soft-mode driven proper ferroelectric phase transition would be unrealistically high.

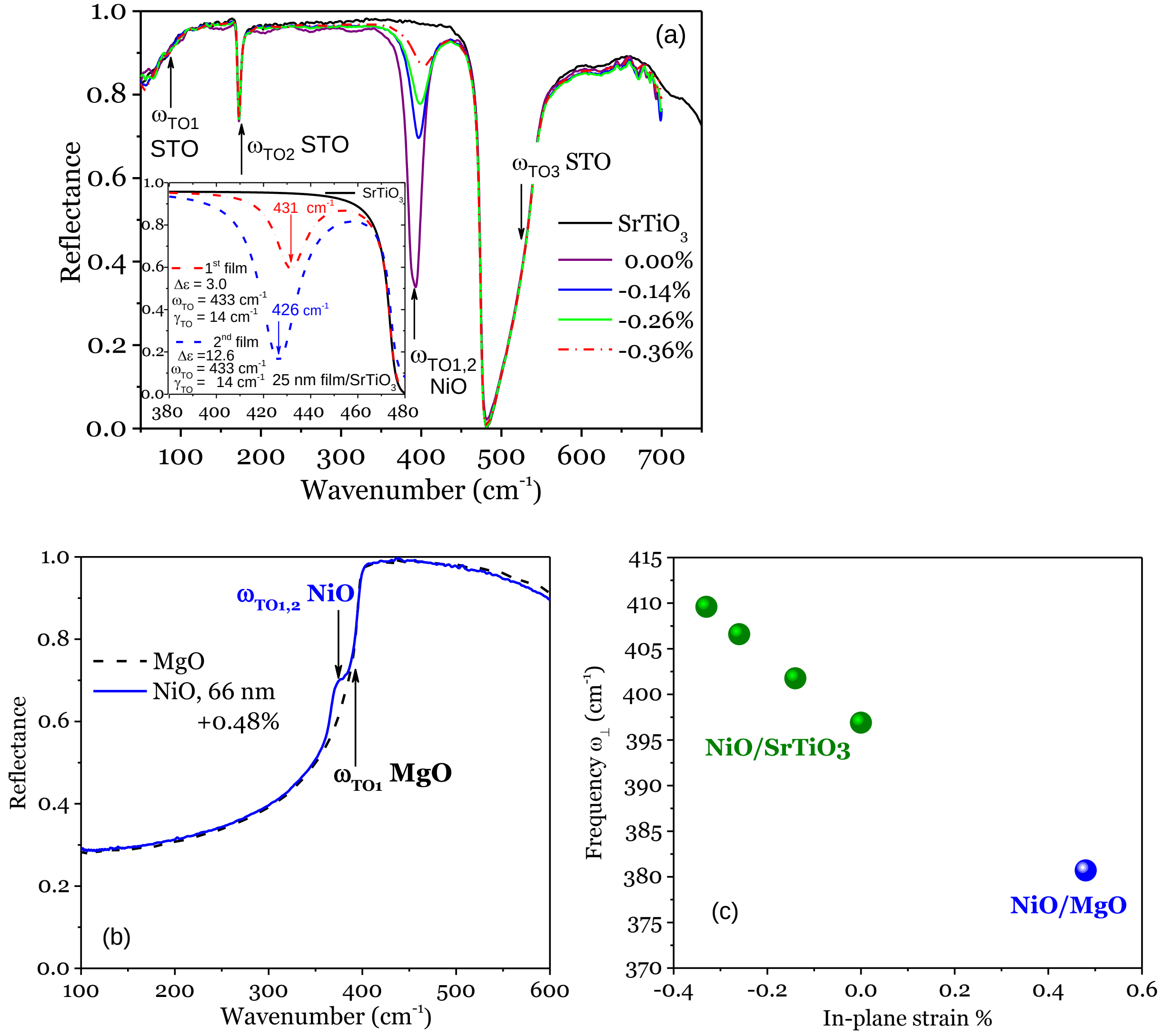

**Fig. 6.** Room temperature IR reflectance spectra of NiO thin films with various strains grown on (a) (001) STO and (b) (001) MgO substrates. (c) The room-temperature $\omega_\perp$ frequency of the TO phonon versus strain. Inset of (a) demonstrates the influence of dielectric strength Δε of the mode in the fictive thin film on frequency of reflectance minima (model phonon parameters are seen in the inset).

Both phonons ($\omega_\parallel$ and $\omega_\perp$) become more pronounced on cooling due to the reduced phonon dampings (Fig. 7). Decrease of the temperature was accompanied by the increase (hardening) of the frequency of both modes $\omega_\parallel$ and $\omega_\perp$ in all deposited films with different levels of strain (insets in fig. 7). Phonons generally harden on cooling due to the anharmonic effects (lattice contraction etc.) and their frequencies $\omega_{TO}$ have a tendency to saturation at low temperatures. This temperature dependence can be in non-magnetic materials described by the formula [33].

$$\omega_{TO}(T) = \omega_0 \left( 1 - \frac{c}{\exp\left(\frac{\theta_D}{T}\right) - 1} \right) \tag{6}$$

where $\omega_0$ indicates the eigenfrequency at 0 K, $c$ is a mode-dependent scaling factor and $\theta_D$ marks the Debye temperature. Nevertheless, we did not use Eq. (6) for fits of the temperature dependence of NiO phonon frequencies, because they are influenced by the spin-phonon coupling, which can violate the dependence in Eq. (6).

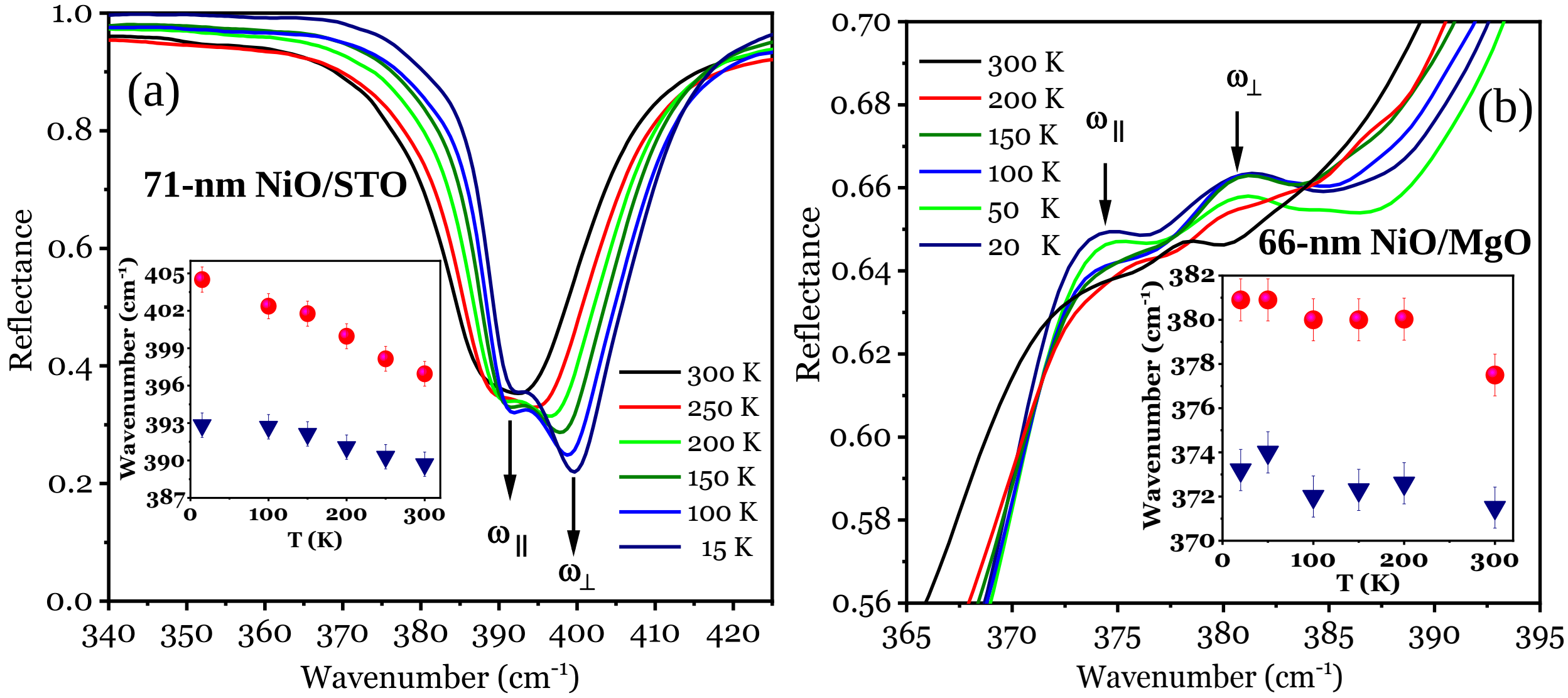

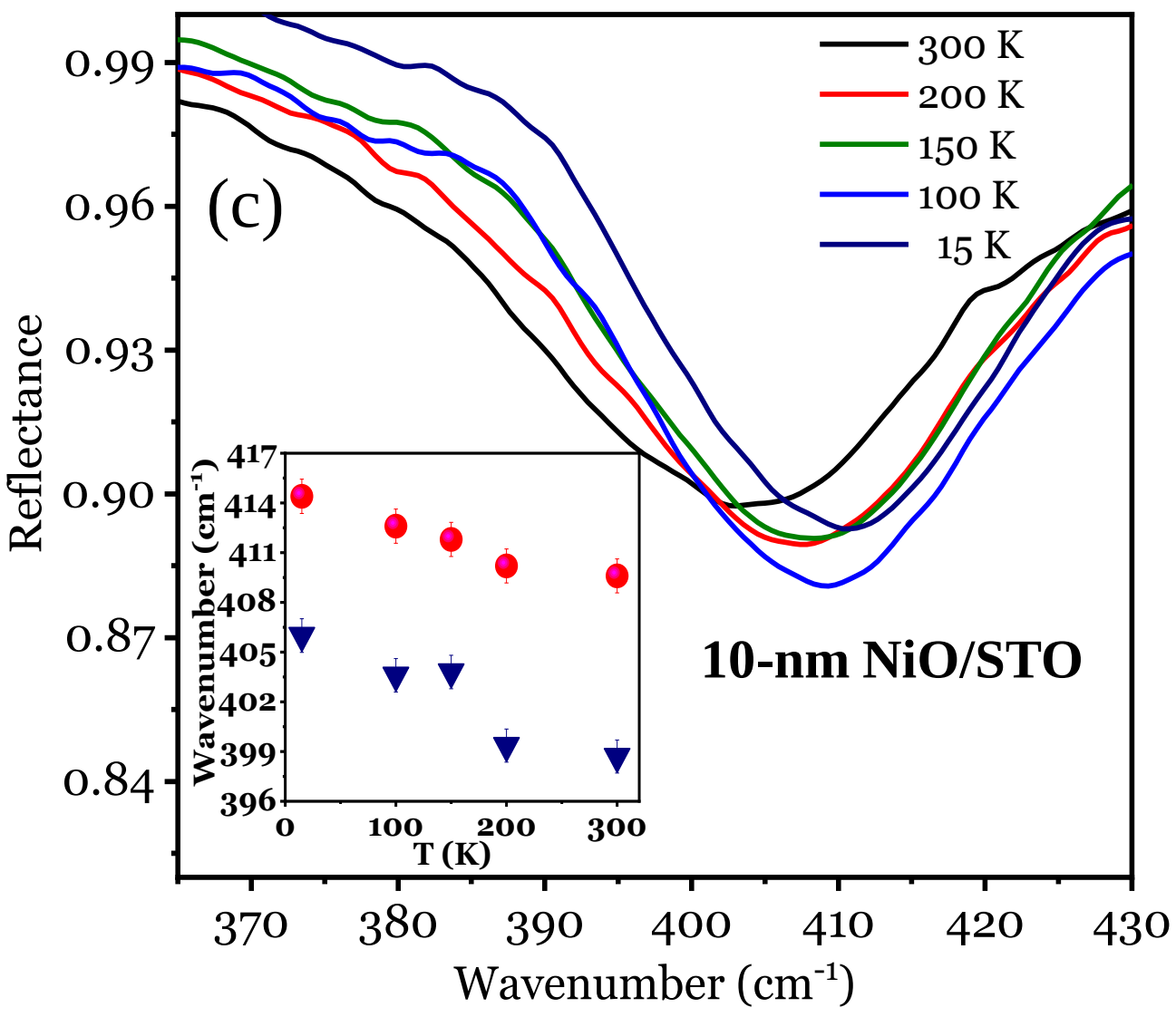

**Fig. 7.** IR reflectance spectra of (a) 71-nm (relaxed) and (b) 66-nm (+0.48% strain) (c) 10-nm (-0.36% strain) NiO thin films at different temperatures. Insets show the temperature dependence of frequencies $\omega_{\parallel}$ (blue) and $\omega_{\perp}$(red).

**Table. 2.** The frequencies of lower $\omega_{\parallel}$ and upper $\omega_{\perp}$ phonon frequencies in NiO thin films with different level of strain measured at ~15 K.

| Sample | In-plane strain | $\omega_{\parallel}$ ($cm^{-1}$) at ~ 15 K | $\omega_{\perp}$ ($cm^{-1}$) at ~ 15 K |
|---|---|---|---|
| 66-nm NiO on MgO | 0.48% | 373.2 | 380.9 |
| 71-nm NiO on STO | 0.0% | 392.8 | 404.5 |
| 10-nm NiO on STO | -0.36% | 406.0 | 414.4 |

Table 2 lists all phonon frequencies obtained from the fits of 15-K infrared spectra of three samples with various strains. Both $\omega_{\parallel}$ and $\omega_{\perp}$ phonon frequencies harden almost 33 $cm^{-1}$ as a result of strain variation from 0.48% to -0.36%. Fig. 8 shows the size of splitting, i.e.$|\Delta\omega|$, as a function of temperature in the samples with different levels of strain. In the relaxed film, $|\Delta\omega|$ increased from 7.3 $cm^{-1}$ (at RT) to 11.7 $cm^{-1}$ (at 15 K) as temperature decreases. This behavior is similar to the bulk NiO. At 15 K, the size of the splitting is slightly larger than that in the bulk NiO [4, 22], but still within the accuracy of the measurements.

In 0.48% tensile strained NiO film, the phonon splitting $|\Delta\omega|$ also increases on cooling, but the splitting is systematically 3 - 4 $cm^{-1}$ smaller than that in the relaxed NiO film or NiO crystal. Since $|\Delta\omega|$ is completely determined by $|J_1|$ [18, 22], the $|J_1|$ must be reduced by the tensile strain which is actually seen in theoretical fig. 2(a). Most peculiar behavior exhibits phonon splitting in the 0.36% compressively strained film. At RT $|\Delta\omega|$ is 4 $cm^{-1}$ larger than that in the relaxed film and it reduces on cooling such that, at 15 K, it becomes smaller than that in relaxed film. It can indicate the decrease of $|J_1|$ on cooling in compressively strained film, but a more plausible explanation is just higher inaccuracy of the mode parameters determination from the fig. 7(c), because the compressively strained film was only 10 nm thin and the obtained phonon signal was much weaker than that in the other two thicker films. For that reason, we do not want to speculate here about the influence of the compressive strain on $J_1$ coupling constant.

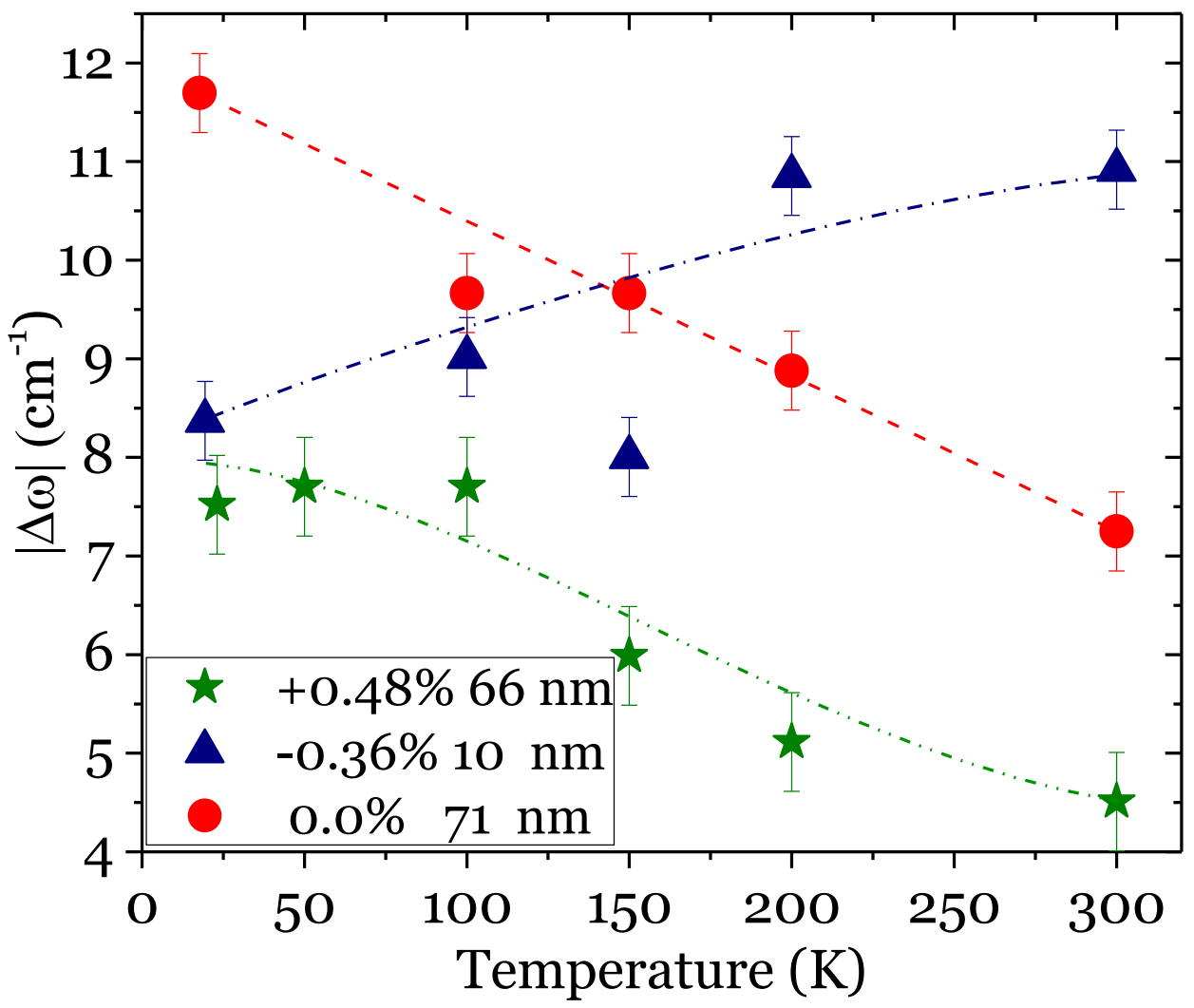

**Fig. 8.** The value of exchange-driven phonon splitting $|\Delta\omega|$ as a function of temperature for three NiO films with different levels of the strain. Lines are guides for eyes.

We recently investigated the influence of strain on the phonon splitting in MnO films grown on MgO substrates, where the phonon splitting increased 20% with ~ 0.2 % compressive strain [14]. The important difference between NiO and MnO is the sign and size of nearest neighbor coupling constant $J_1$. NiO has a small and negative $J_1$ and; therefore, a negative $\Delta\omega = \omega_{||} - \omega_{\perp}$, while MnO has $J_1$ much larger and positive $\Delta\omega$ [19]. According to Kant *et al.* [22], the energy of phonon splitting is linearly dependent on $J_1$ and follows the relation $\hbar\Delta\omega = \beta J_1 S^2$. It explains negative $\Delta\omega$ in NiO and positive $\Delta\omega$ in MnO.

Fischer *et al.* [19] calculated $J_1$ in various monoxides as a function of lattice parameters and found that $J_1$ is only slightly dependent on the lattice parameter in NiO, but strongly dependent in MnO (Fig. 2). For that reason, the change of $|\Delta\omega|$ with the strain is larger in MnO than in NiO. In our case, we observed a reduction of phonon splitting by 3 – 4 $cm^{-1}$ with 0.48 % tensile strain in NiO, while $|\Delta\omega|$ increases by 5 $cm^{-1}$ with 0.2% compressive strain in MnO. It qualitatively confirms the theoretical predictions that $|\Delta\omega|$ is solely determined by $J_1$.

## Conclusion

NiO (001) thin films with various strains were grown on $SrTiO_3$ (001) and MgO (001) substrates. The films grown on the STO substrate showed an in-plane compressive strain, which increased as the film thicknesses were reduced, resulting in a smooth increase (hardening) of the phonon frequencies. On the other hand, a tensile strain was detected in the NiO films grown on the MgO substrate inducing a strong softening of the phonons. The magnetic exchange-driven phonon splitting $\Delta\omega$ in three different films with 0.36% compressive, 0.48% tensile strain and the film with zero strain, were measured as a function of temperature. The $|\Delta\omega|$ increased on cooling in NiO tensile strained and relaxed films. Compressively strained NiO film exhibits a small decrease of $|\Delta\omega|$ on cooling, but this observation can be an artifact of less accurately determined phonon frequencies in 10 nm thin film. The phonon splitting is systematically 3 – 4 $cm^{-1}$ lower in the tensile strained film than that in the bulk NiO or in NiO film with relaxed strain. It gives a hint that $|J_1|$ reduces with tensile strain in NiO as it was predicted in [19]. On other hand the phonon splitting increases with compressive strain in MnO thin film, which supports theoretical predictions [19, 22] that $\Delta\omega \propto J_1$ and $J_1$ strongly depends on lattice parameter in MnO. The observed sensitivity of $J_1$ on the strain is highly promising for the study of highly strained MnO films, where spin-order-induced multiferroic phase was predicted [16].

## Acknowledgements

This research has been supported by TJ Park Science Fellowship of POSCO TJ Park Foundation, by the Czech Science Foundation (Project No. 18-09265S) and by Operational Programme Research, Development and Education (financed by European Structural and Investment Funds and by the Czech Ministry of Education, Youth and Sports), Project No. SOLID21 - CZ.02.1.01/0.0/0.0/16_019/0000760.